\RequirePackage[hyphens]{url}

\documentclass[]{interact}
\usepackage[utf8]{inputenc}
\usepackage{dblfloatfix}
\usepackage{amsmath}
\usepackage{braket}
\usepackage{graphicx}
\usepackage{subcaption}
\usepackage[version=4]{mhchem}
\usepackage{comment}
\usepackage{fixmath}
\usepackage{float}
\usepackage{xcolor}
\usepackage[numbers,sort&compress,merge]{natbib}
\usepackage[]{hyperref}

\newcommand{\td}{\text{d}}
\newcommand{\der}[2]{\frac{\td#1}{\td#2}}
\newcommand{\dd}[1]{\td#1}

\begin{document}
\title{Laser Excitation of Muonic 1S Hydrogen Hyperfine Transition: \\Effects of Multi-pass Cell Interference}

\author{
\name{M.~Ferro\textsuperscript{1,2,a}\thanks{\textsuperscript{a}Corresponding authors: \href{mailto:mc.ferro@campus.fct.unl.pt}{mc.ferro@campus.fct.unl.pt} and \href{mailto:pdamaro@fct.unl.pt}{pdamaro@fct.unl.pt}}, 
P.~Amaro\textsuperscript{1,a}, 
L.~Sustelo\textsuperscript{1},
L.~M.~P.~Fernandes\textsuperscript{3},
E.~L. Gründeman\textsuperscript{4}, 
M.~Guerra\textsuperscript{1},
C.~A.~O.~Henriques\textsuperscript{3},
M. Kilinc\textsuperscript{4},
K.~Kirch\textsuperscript{2,4},
J.~Machado\textsuperscript{1},
M.~Marszalek\textsuperscript{2,4},
J.~P.~Santos\textsuperscript{1},
A.~Antognini\textsuperscript{2,4}}
\affil{
\textsuperscript{1}Laboratory of Instrumentation, Biomedical Engineering and Radiation Physics (LIBPhys-UNL), Department of Physics, NOVA School of Science and Technology, NOVA University Lisbon, 2829-516 Caparica, Portugal \\[0.5em]
\textsuperscript{2}PSI Center for Neutron and Muon Sciences, 5232 Villigen, Switzerland \\[0.5em]
\textsuperscript{3}LIBPhys-UC, Department of Physics, University of Coimbra, 3004-516 Coimbra, Portugal \\[0.5em]
\textsuperscript{4}Institute for Particle Physics and Astrophysics, ETH Zurich, 8093 Zurich, Switzerland
}}
\maketitle

\begin{abstract}

Calculating the laser-induced transition probability by using the fluence distribution that neglects wave-interference effects (e.g., by employing ray-tracing methods) can lead to an 
overestimation of this probability, as it underestimates saturation effects. In this paper, we investigate how interference effects in 
the multi-pass cell, used to enhance the laser fluence, affect the laser-induced transition 
probability between hyperfine levels in muonic hydrogen, a bound system of a negative muon and a proton. To avoid complications related to the 
exact knowledge of the intra-cavity field, we develop a simple model that estimates the maximal 
possible interference effects for given laser and multi-pass cell parameters, thereby providing 
an upper bound for the resulting decrease in transition probability relative to the case where 
these effects are neglected. A numerical evaluation of this upper bound for muonic hydrogen 
shows that, under our experimental conditions, such effects can be safely neglected.
Nonetheless, the methodology presented here could be applied to estimate the impact of 
interference effects on the laser-induced transition probability in other experiments 
involving coherent light in multi-pass systems.

\end{abstract}


\section{Introduction}

The proton charge radius is a key benchmark for testing our understanding of proton structure and sets the ultimate precision limit in comparing theory and experiment in simple atoms  and molecules such as H, H$_2^+$ and HD$^+$~\cite{KARSHENBOIM20051, IEIDES200163, antognini_2022}. Its determination took center stage with the emergence of the proton radius puzzle~\cite{pohl2013muonic, peset2021proton, karr2020proton, gao2022proton}, sparked by laser spectroscopy of the 2S-2P in muonic hydrogen ($\mu$p), which revealed a striking discrepancy with values from electronic hydrogen and electron–proton scattering~\cite{pohl_2010,antognini_2013,pohl_2016}. The finding ignited a wave of cross-disciplinary activity—ranging from new laser spectroscopy experiments~\cite{grinin_2020, beyer_2017, Brandt:2021yor, Bezginov2019, Scheidegger_PhysRevLett.132.113001} and scattering experiments~\cite{Xiong2019, gilman2017technicaldesignreportpaul, Lin:2021umz, AMBER} to precision spectroscopy of simple molecular ions and molecules~\cite{Patra, Alighanbari2020, Holzapfel_PhysRevX.15.031009, Alighanbari2025, holsch2019benchmarking}, advances in bound-state QED~\cite{karshenboim_2015, Pohl_RevModPhys.96.015001, Adkins:2024wws}, refined descriptions of hadron structure via chiral perturbation theory, dispersion relations, and lattice QCD~\cite{Lin:2021umz, peset2021proton, HAGELSTEIN2021122323, Alexandrou:2025vto} and beyond standard model searches~\cite{carlson2014constraints, Potvliege:2024xly}. Fifteen years on, most new measurements and re-analyses now agree with the  value from $\mu$p, transforming the puzzle into a precision benchmark~\cite{antognini_2022}.

Building on the pivotal role of the proton charge radius as both a discriminator between theoretical approaches and a catalyst for progress across diverse areas of physics,
the CREMA collaboration is now aiming to measure the ground-state hyperfine splitting (HFS) ~\cite{Nuber2023, amaro_2022} in $\mu$p to address the magnetic properties of the proton~\cite{ruth2024new, antognini_2022, Djukanovic_PhysRevD.110.L011503, Hagelstein2023, LIN2024419}. 
Furthermore, combining the $\mu$p HFS values (both theoretical and experimental) with the corresponding HFS measurements in H~\cite{ESSEN1971, Bullis_PhysRevLett.130.203001} allows for a stringent test of our understanding of hyperfine splitting effects, including challenging higher-order corrections~\cite{IEIDES200163, Pachucki:2024hwi, Pachucki:2025kde}, and provides sensitivity to potential physics beyond the Standard Model~\cite{antognini_2022, Frugiuele2022}.

The transition under investigation is a magnetic dipole (M1) transition and therefore electric-dipole forbidden, resulting in a low intrinsic transition probability and consequently a limited signal rate. Accurate quantification of the laser-induced transition probability is therefore critical. To enhance the excitation probability, the laser pulses are injected into a multi-pass cell, which significantly increases the laser fluence within the interaction volume. Ray-tracing simulations of multi-pass cells can be used to estimate the spatial distributions and average laser fluences~\cite{mirek_fluence}; however, they inherently neglect wave-interference effects between multiple passes. Such interference can lead to spatial modulation of the intensity distribution and thus enhance saturation effects. In this work, we provide an upper bound on the reduction of the laser-induced transition probability attributable to interference effects in the multi-pass cell used in the measurement of the HFS in $\mu$p. 
This analysis can be used to inform cavity design optimization, mitigating adverse interference-related saturation effects, and to refine the signal rate estimation in the experiment.

In Sec.~\ref{sec:exp_scheme}, we summarize the experimental principle of the CREMA HFS experiment needed to understand the conditions in which laser excitation is taking place. In Sec.~\ref{sec:theory}, we present our approach to compute the laser-induced  transition probability. We begin by describing the multi-pass cell and how we simplify the light propagation within it to extract an upper bound for the interference effects and the associated reduction of the laser-induced transition probability. We then present the mathematical description of the electric field in the cell corresponding to this worst-case scenario and how it is incorporated into the optical Bloch equations. 
Section~\ref{sec:Monte-Carlo} describes the numerical implementation.
The results—comprising the fluence distributions from stochastic interference effects and the corresponding reduction in excitation probability under various conditions—are presented in Sec.~\ref{sec:results}.

\section{Principle of the HFS measurement in $\mu$p}
\label{sec:exp_scheme}

\begin{figure}[tb]
	\centering
    \includegraphics[width=0.75\textwidth]{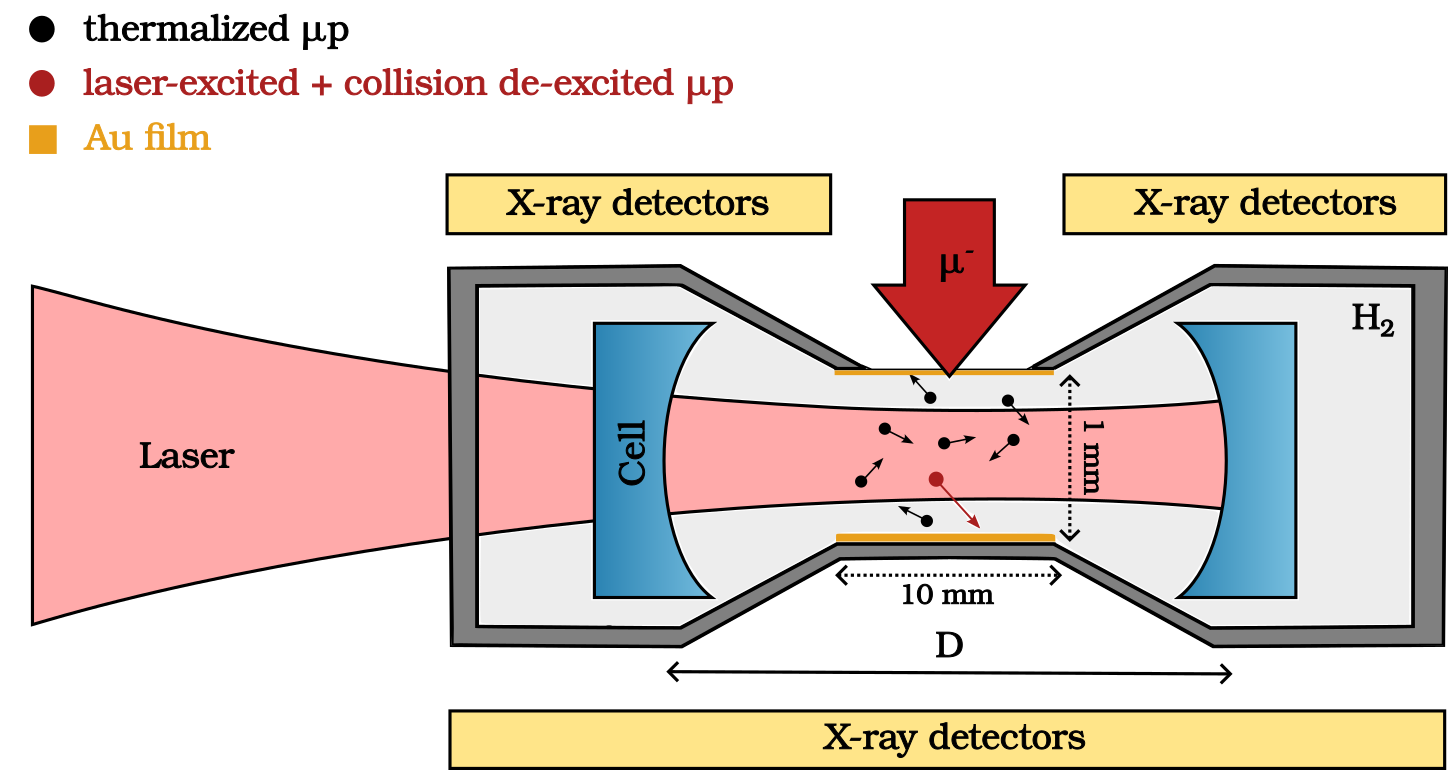}
	\caption{Schematic (not to scale) of the experimental setup for the CREMA HFS experiment. The dashed lines represent the dimensions of the muon stopping volume and the solid line represents the cell's diameter, $D$.}
    \label{fig:scheme}
\end{figure}

\begin{figure}[b!]
\centering
\includegraphics[width=0.9\textwidth]{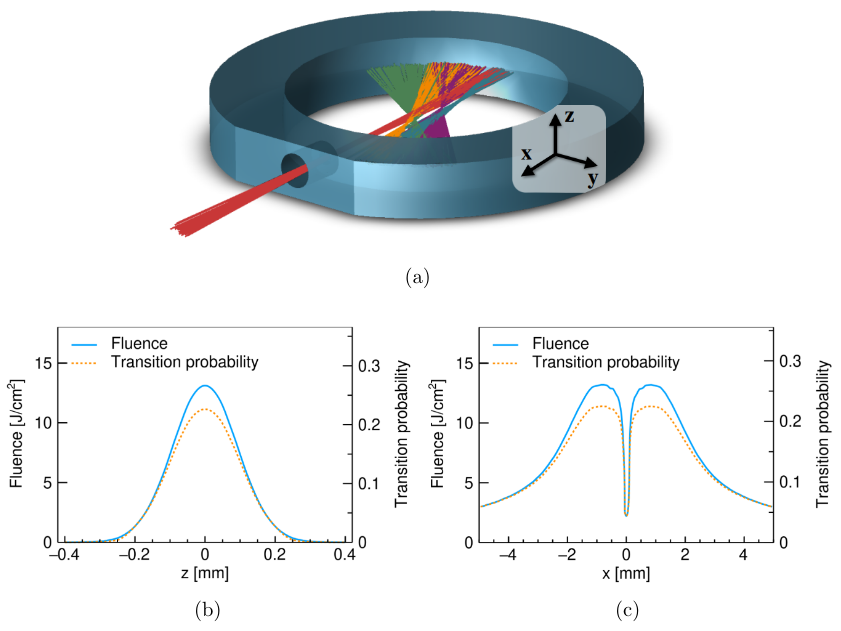}
\caption{(a) 3D rendering of the toroidal multi-pass cell with an example of a ray-tracing simulation. The laser pulse enters through a 0.5 mm-thick slit at a small angle relative to the x-axis. To illustrate the beam path, the first few reflections inside the cell are shown in a different colour. (b) Spatial distribution of fluence along the $z$-axis (muon beam axis) of the toroidal cell, obtained from ray-tracing simulations, and respective transition probability (laser excitation followed by collisional de-excitation) for an in-coupled pulse of 1~mJ energy and $R=0.992$. The transition probability is calculated for a $\mu$p atom in the $F = 0$ state assuming a
laser bandwidth of 100~MHz, a pulse length of 50~ns, a target temperature of 22~K and
a target pressure of 0.6~bar. (c) Same as (b) but for the $x$-axis. Figure and caption reproduced from~\cite{Nuber2023}. 
} \label{fig:cavity}
\end{figure}

The principle of the experiment is represented in Fig.~\ref{fig:scheme} and detailed in Ref.~\cite{Nuber2023}. A low energy muon beam of about 11~MeV/c momentum and average rate of 500~s$^{-1}$ is directed towards the H$_2$ target, triggering the laser system upon passing an entrance detector. A fraction of the muons is stopped in a cryogenic H$_2$ gas target at a temperature of 22~K and a pressure of about 0.5~bar leading to the formation of $\mu$p atoms in highly excited states. The atoms quickly de-excite to the 1S state and reach the $F=0$ (singlet) sublevel of the ground state through collisions with the H$_2$ gas. At these target conditions, after about 1~$\mu$s the $\mu$p atoms are thermalized to the H$_2$ gas temperature.  A laser pulse of a few millijoules at a wavelength of 6.8~$\mu$m (corresponding to 44~THz or 0.2\,eV photon energy) is then coupled into a toroidal multi-pass cell (Fig.~\ref{fig:cavity}) illuminating a significant fraction of the muon stopping volume.

The multiple reflections inside the multi-pass cell create a disk-shaped illuminated volume, enhancing the laser fluence at the position of the \(\mu\)p atoms, which are distributed homogeneously around the muon beam axis with a diameter of about 10~mm. The resonant laser light drives the transition of the \(\mu\)p atoms from the singlet \(F=0\) to the triplet \(F=1\) sublevel. The \(\mu\)p atoms in the \(F=1\) excited state collide inelastically with H$_2$ gas molecules within nanoseconds, de-exciting back to the $F=0$ level. During this de-excitation process, the HFS  transition energy is converted into kinetic energy, causing the \(\mu\)p atoms to gain, on average, approximately 0.1~eV of kinetic energy. Since this kinetic energy is much greater than the thermal energy, the de-excited \(\mu\)p atoms diffuse away from the muon stopping volume and, with considerable probability, reach the target walls coated with a thin layer of gold. The negative muon from the \(\mu\)H atoms is transferred to a gold nucleus, forming muonic gold (\(\mu\)Au$^*$) in highly excited states. These muonic gold atoms promptly de-excite through a cascade of radiative transitions, emitting a few X-rays with MeV energies, which serve as a detectable signature of successful laser excitation. The HFS resonance is thus obtained by counting the \(\mu\)Au$^*$ cascade events (within a specified time window) following the laser excitation as a function of the laser frequency.

\section{Theoretical model}
\label{sec:theory}

Key physical processes affecting the laser excitation probability between the 1S hyperfine sublevels include decoherence arising from elastic and inelastic collisions between $\mu$p atoms and H$_2$ molecules, as well as the laser bandwidth. Here, elastic and inelastic refer to collisions that do not change, or do change, respectively, the internal state of the $\mu$p atom. These effects, together with the Doppler effect, were taken into account in the computation of the laser-induced transition probabilities in  Ref.~\cite{amaro_2022}, assuming a flat-top laser pulse.
These probabilities have been integrated into a comprehensive simulation of the experimental setup, encompassing also $\mu$p formation, thermalization, and diffusion in the gas~\cite{Nuber2023}. 
Additionally in this study, the laser excitation probability was evaluated using the spatial fluence distribution in the multi-pass cell, as obtained from ray-tracing simulations, as shown in Fig.~\ref{fig:cavity}. The small deviation in shape between the laser transition probability and the fluence distributions in panels (b) and (c) of this figure arises due to saturation effects in the region where the fluence is highest.
In the present study, we focus on quantifying the impact of wave-interference effects between the multiple passes in the cell, which cannot be captured by ray-tracing simulations. These interference effects lead to additional saturation, further reducing the laser excitation probability.
The results of this investigation are used to optimize the multi-pass cell design and to establish an upper bound for the reduction in laser transition probability (followed by collisional de-excitation) caused by interference-induced saturation effects.

\subsection{Multi-pass cell}
\label{sec:cavity}

The toroidal multi-pass cell for the HFS experiment, shown in Fig.~\ref{fig:cavity}~(a), is stable in the vertical direction ($z$) and unstable in the horizontal plane ($x$, $y$)~\cite{mirek_phd}. 
In the vertical direction, the laser pulse mode is resonant, with a waist at the center of the cell of about $0.2$~mm. This waist is small enough to avoid diffraction losses at the target walls (spaced by 1 mm), while being as large as possible to maximize the overlap with the muon stopping distribution and to minimize saturation effects.

The 50~ns long laser pulses are injected into the cell through a tiny slit  aligned in the $z$-direction. While reflecting off the mirror surfaces, the light gradually spreads within the $xy$-plane, eventually filling a disk-shaped volume, thereby enhancing the laser fluence within this region. 
The spatial length of the pulse (approximatively 15 m) is much larger than the diameter $D=10$~cm of the cell, causing successive reflections to interfere with each other.
\begin{figure}[b!]
    \centering
    \includegraphics[width=0.65\textwidth]{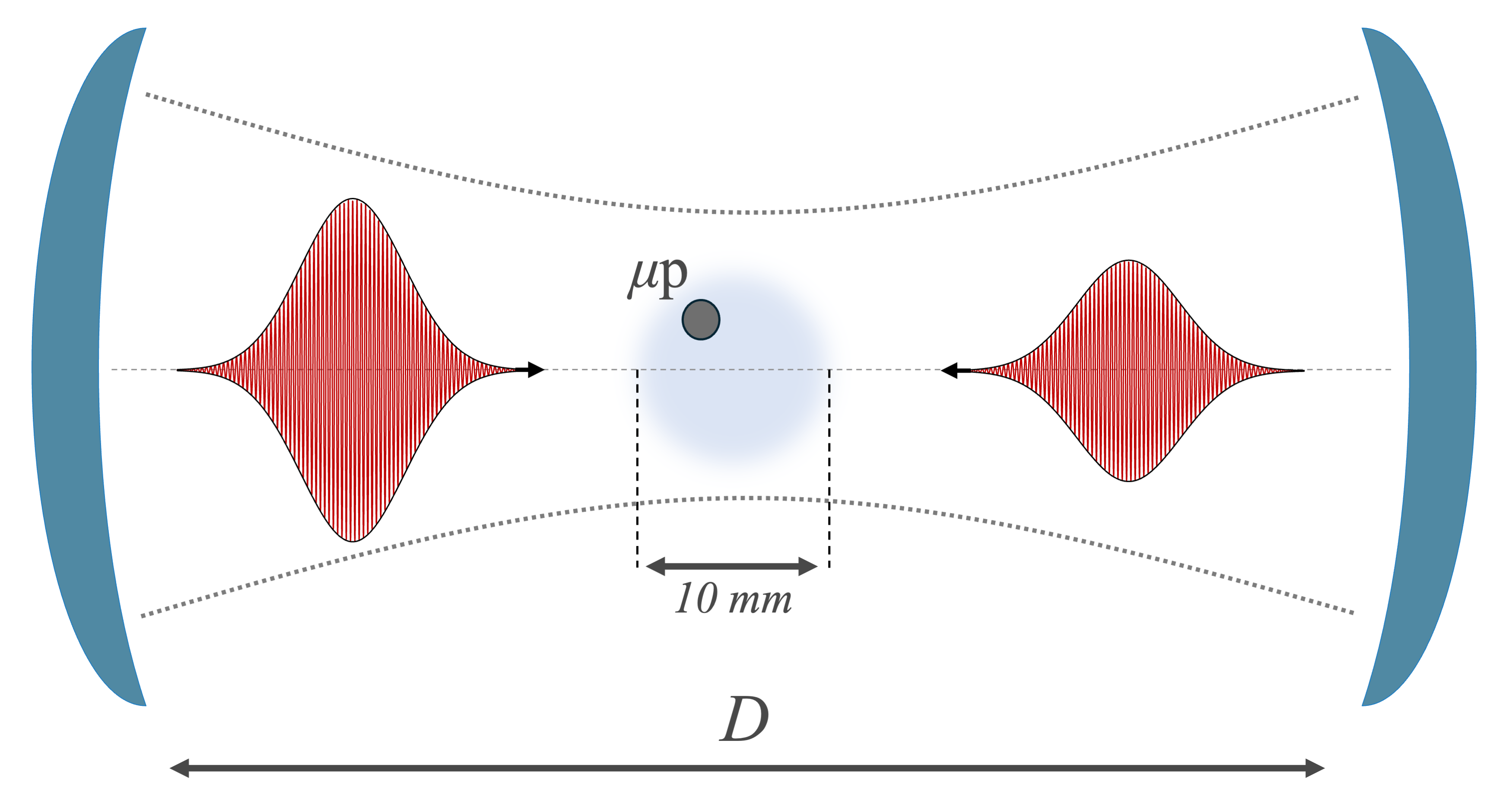}
	\caption{Simplified model of the multi-pass cell used to estimate the maximal possible interference effects. The model is one-dimensional, consisting of two reflecting surfaces with  reflectivity $R$ separated by a distance $D$. A laser pulse undergoes multiple reflections between the surfaces, interfering with itself. The pulse length is not drawn to scale; in the cases relevant to this paper and for the considered HFS experiment, its spatial extent far exceeds the mirror separation $D$. The $\mu$p location is indicated by the blue shaded region.}
	\label{fig:toroidal_cavity}
\end{figure}

To investigate the maximal saturation effects arising from interference between the laser pulse and its reflections within the multi-pass cell, we simplify the light propagation to a one-dimensional model, as illustrated in Fig.~\ref{fig:toroidal_cavity}. This model consists of two parallel reflecting surfaces separated by a distance $D$, each with (intensity) reflectivity $R$. The electric field is treated classically as a sum of time-delayed pulses (see Sec.~\ref{sec:elecfi}), capturing interference between multiple reflections. This resulting field then drives the time-dependent optical Bloch equations (Sec.~\ref{sec:levelpopo}).

To maximize interference effects, we assume linear polarization oriented along the vertical ($z$) direction. In contrast, linear polarization in the $xy$-plane would produce reduced interference because the polarization vector rotates with each reflection, causing partially orthogonal polarization states between interfering beams.

Because the spatial fluence distribution in the cell depends strongly on input beam parameters such as angle and divergence, in this study we investigate the interference effects as a function of the average fluence and do not consider any detailed spatial dependence of the electric field.

In our model, to account for the varying path length at each reflection (i.e., the distance from the $\mu$p atom to the mirror and back), we assume uncorrelated random phase delays for successive passes of the folded beam impinging on the $\mu$p atom.
Since the overlap of the various beams in the real 3D geometry is smaller than in the simplified 1D model, the latter overestimates the overlap between the original pulse and its reflections, thereby exaggerating the interference effects. Consequently, this model yields a conservative upper bound on interference-induced saturation effects in the toroidal multi-pass cell.

\subsection{Electric field and fluence models}
\label{sec:elecfi}

Even though this paper deals with a magnetic M1 transition, we model the B-field using the more commonly used electric field and then relate the two via $E=cB$.
The electric field inside the laser multi-pass cell is modelled as an initial pulse that is successively reflected at the reflecting surfaces, interfering with itself. The time evolution of the electric field strength  at the $\mu$p position can be thus represented by the sum of  all reflections,
\begin{equation}
\label{eq:field}
	E_k(t) = E_L\sum_{n=0}^{\infty} (\sqrt{R})^nG_{t_n, \tau}(t) \cos(\omega t + \phi_{k,n}) ,
\end{equation}
\noindent where $E_L$ is the field's amplitude of the incoupled laser pulse, $\omega$ the pulse's angular frequency, $\phi_{k,n}$ the phase of the $n$-th reflection, and 
\begin{equation}
\label{eq:gauss}
	G_{t_n, \tau}(t) = \left(\frac{1}{\tau\sqrt{\pi}}\right)^{1/2}\exp\left[-\frac{(t-t_n)^2}{2\tau^2}\right]
\end{equation}
\noindent 
the Gaussian amplitude of the laser pulse with temporal width $\tau$, centered at time $t_n = t_0 + n t_r$. Here, $t_0$ denotes the time when the pulse is coupled into the cell. If the atom is in the center of the cell, $t_r = D/c$ is the time required for the pulse to travel back and forth between the atom and the cell surface upon reflection. If the atom is displaced from the center, the same equations hold but $t_r=D/c$ represents the average time for back and forth propagation between the atom and the cell surface. The function $G_{t_n, \tau}(t)$ is normalized such that
\begin{equation}
\label{eq:gauss_norm}
	\int_0^{\infty} G_{t_n, \tau}^2(t) \dd{t} = 1.
\end{equation}
The factor $(\sqrt{R})^n$ accounts for the reduction in amplitude at each reflection, where the amplitude decreases by a factor of $\sqrt{R}$ per reflection.
 As explained above we consider that each reflected pulse has its phase shifted by a random amount, i.e., each phase $\phi_{k,n}$ in Eq.~\eqref{eq:field} is sampled from a uniform distribution from 0 to $2\pi$. The index $k$ addresses a unique stochastic generation of the total field in the cavity, for the $k$-th injected pulse.

The fluence~\cite{mirek_fluence} associated to the $k$-th injected pulse, is given by
\begin{equation}
	\mathcal{F}_k = \varepsilon_0c\int_{0}^{\infty} |E_k(t)|^2 \dd{t}.
\end{equation}
\noindent Substituting the electric field of Eq.~\eqref{eq:field}, the fluence becomes
\begin{equation}
	\label{eq:fluence_total}
	\mathcal{F}_k = \frac{\mathcal{F}_0}{1-R} + (\mathcal{F}_{\text{int}})_k,
\end{equation}

\noindent where $\mathcal{F}_0$ is the fluence of the laser pulse injected into  the multi-pass cell:
\begin{equation}
	\mathcal{F}_0 = \varepsilon_0 c \int_{0}^{\infty} |E_LG_{t_0, \tau}(t)\cos(\omega t + \phi_0)|^2 \dd{t} = \frac{\varepsilon_0 c E_L^2}{2},
\end{equation}

\noindent assuming $(\omega\tau)^2 \gg 1$. 

The first term in Eq.~\eqref{eq:fluence_total} represents the well known cavity enhancement factor while  the second term represents the correction that accounts for the interference between the various  reflections. The latter  can be expressed as:
\begin{equation}
	\label{eq:fluence_interf}
	\begin{aligned}
		(\mathcal{F}_{\text{int}})_k = \varepsilon_0cE_L^2 \sum_{i>j}(\sqrt{R})^{i+j}&\exp\left[-\frac{t_r^2(i-j)^2}{4\tau^2}\right] 
		\times\cos(\phi_{k,i}-\phi_{k,j}) .
	\end{aligned}
\end{equation}
The average fluence, over many injected pulses ($k = 0, 1, ..., K$) for the same values of $\mathcal{F}_0$, $\tau$, $D$ and $R$, reduces to the well-known result 
\begin{equation}
	\label{eq:fluence_avg}
	\overline{\mathcal{F}} = \frac{\mathcal{F}_0}{1-R} + \overline{\mathcal{F}_{\text{int}}} = \frac{\mathcal{F}_0}{1-R},
\end{equation}

\noindent as the average of the interference term, $\overline{\mathcal{F}_{\text{int}}}$, is zero. For this reason, the interference is typically neglected. However, when accounting for saturation effects in the atoms' transition probability, the interference term cannot be neglected as the average transition probability depends on the distribution of $\mathcal{F}_{\text{int}}$  for fixed  $\overline{\mathcal{F}}$, $\tau$, $D$ and $R$.

Numerical results of the distribution of $\mathcal{F}_k$ are presented in Sec.~\ref{sec:results}. These distributions do not include any position dependence, since the electric field of each individual pass is position independent. Any position-dependence of the fluence in the 3D cell can be accounted for by analysing the interference effect as a function of the average fluence.

\subsection{Optical Bloch equations and level's populations}
\label{sec:levelpopo}

\begin{figure}[htb]
    \centering
    \includegraphics[width=0.75\textwidth]{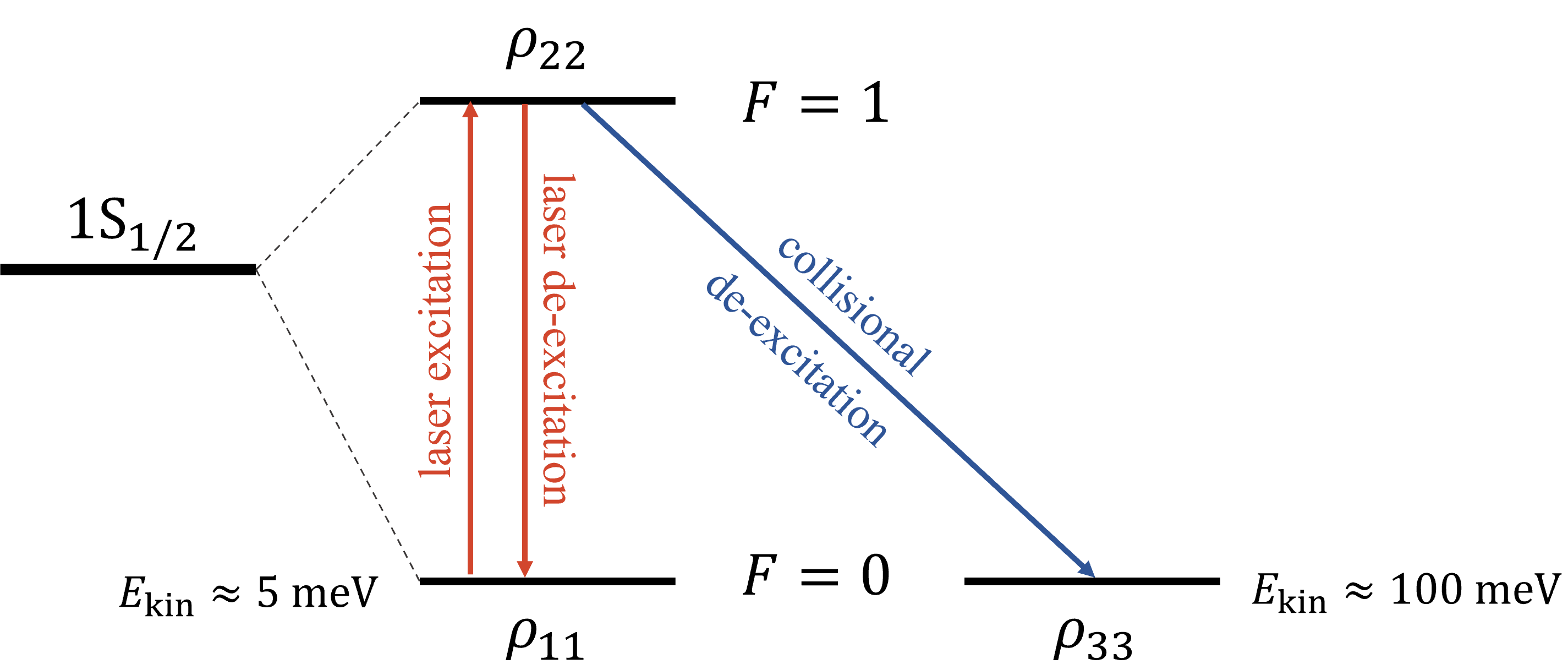}
	\caption{
   Diagram of the HFS sub-levels showing the laser-induced transition and the collisional de-excitation, which produces a $\mu$p atom with approximately 100\, meV of kinetic energy~\cite{amaro_2022}. 5~meV corresponds to the kinetic energy of $\mu$p atoms thermalized at a temperature of 22\,K.}
    \label{fig:energy_levels}
\end{figure}
We compute the laser-induced transition probability and the corresponding time evolution of the sublevel population densities (see Fig.~\ref{fig:energy_levels}) using the density-matrix formalism and  explicitly including both elastic and inelastic collisional effects. 
The dynamics are governed by the optical Bloch equations, which in this context read~\cite{amaro_2022}:
\begin{equation} \label{eq:bloch_equations}
	\begin{aligned}
		\der{\rho_{11}}{t}(t) &= -\operatorname{Im}\left(\Omega(t)\rho_{12}e^{i\Delta\omega t}\right) + \Gamma_{\text{sp}}\rho_{22},\\[1em]
		\der{\rho_{22}}{t}(t) &= \operatorname{Im}\left(\Omega(t)\rho_{12}e^{i\Delta\omega t}\right) - (\Gamma_i + \Gamma_{\text{sp}})\rho_{22},\\[1em]
		\der{\rho_{12}}{t}(t) &= i\frac{\Omega^*(t)}{2}(\rho_{11} - \rho_{22})e^{-i\Delta\omega t} - \frac{\Gamma_c}{2}\rho_{12},\\[1em]
		\der{\rho_{33}}{t}(t) &= \Gamma_i\rho_{22}.
	\end{aligned}
\end{equation}

\noindent Here $\rho_{11}$ and $\rho_{22}$  represent the population density of the thermalized $\mu$p atom in the $F=0$ and  $F=1$ sublevels and $\rho_{33}$ represents the  population density of the $F=0$ state right  after  a collisionally de-excitation from the $F=1$ state, so that $\rho_{33}$ can be understood as the combined probability of laser excitation followed by collisional de-excitation.
 $\Omega(t)$ is the Rabi flopping frequency which depends on the strength of the  laser field, and $\Delta\omega = \omega_r - \omega$ is the angular frequency detuning between laser frequency ($\omega$) and resonance frequency ($\omega_r$) for the $F=0 \longrightarrow F=1$ transition. 
 Since we assume the rotating wave approximation, terms oscillating at the sum frequency $\omega_r + \omega$ are neglected. The total decoherence rate~\cite{amaro_2022}
$	\Gamma_c = \Gamma_i + \Gamma_e + \Gamma_{\text{sp}} + 2\pi\Delta_l$ 
in the off-diagonal density matrix elements, accounts for elastic and inelastic collision rates, $\Gamma_e$ and $\Gamma_i$, the spontaneous emission rate $\Gamma_{\text{sp}}$ from $F=1$ to $F=0$, and the laser bandwidth $\Delta_l$ given in Hertz. 
Throughout this work we used $\Gamma_e=72$~MHz, $\Gamma_i=82$~MHz, and $\Gamma_{\text{sp}}=1.23\times 10^{-11}$~MHz, as calculated in ~\cite{amaro_2022} assuming a target at 22~K and 0.5~bar pressure.

It is important to note that the Bloch equations in Eqs.~\eqref{eq:bloch_equations} assume that the third state, described by $\rho_{33}$, is a dark state. This is because, after collisional de-excitation, the muonic hydrogen atom is Doppler-shifted out of resonance due to the acquired kinetic energy. At our target conditions, the subsequent thermalization occurs on the  1~$\mu$s timescale, as shown in \cite{Nuber2023}.
Hence, a direct link between $\rho_{33}$ and $\rho_{11}$ can be safely neglected in this study.

\section{Monte-Carlo implementation}
\label{sec:Monte-Carlo}
We repeatedly numerically integrated the optical Bloch equations to obtain $(\rho_{33})_k$ for stochastically generated electric fields $\mathcal{E}_k$. The integration is performed beyond the duration of the laser pulse to allow the muonic hydrogen in the triplet state to collisionally relax to the ground state, as it occurs in the actual experiment.

These simulations were then performed for a range of laser and multi-pass cell parameters ($\tau, \,D, \, R, \, \overline{\mathcal{F}}$ and $ \, \Delta \omega$), in order to study their influence on the additional saturation effects we aim to quantify.

Practically, a dimensionless electric field $\mathcal{E}_k$, including interference effects according to Eq.~\ref{eq:field}, is constructed by generating $N$ random phases $\phi_{k,n}$:
\begin{equation}
	\label{eq:dimensionless_field}
	\mathcal{E}_k (t)\equiv \frac{E_k(t)}{E_L e^{i \omega t}} = \sum_{n=0}^{N} (\sqrt{R})^nG_{t_n, \tau} e^{i\phi_{k,n}}.
\end{equation}

\noindent 
The number $N$, representing the total number of reflections of the laser pulse considered in the numerical implementation, is chosen such that the amplitude of the last reflected pulse is 1\% of the initial pulse amplitude. 
 This condition gives $N = \ln(0.01)/ \ln(\sqrt{R}) $.
No significant variations of the results   were observed for larger values of $N$. Therefore, to save computation time in the systematic evaluation of the interference effect, we adopted this choice.
Note that, unlike in Eq.~\ref{eq:field}, the normalized electric field $\mathcal{E}_k$ in Eq.~\ref{eq:dimensionless_field}
does not contain the oscillatory term $e^{i\omega t}$, since this factor is already included in each $e^{i \Delta \omega t}$ term of the Bloch equations given in Eqs.~\ref{eq:bloch_equations}.

The corresponding time-dependent Rabi flopping frequencies, to be used in the optical Bloch equations of Eqs.~\eqref{eq:bloch_equations}, can be evaluated from the normalized electric fields $\mathcal{E}_k$, from the fluence $\mathcal{F}_0$ of the in-coupled laser pulse,  and from the matrix element $\mathcal{M}$:
\begin{equation}
	\label{eq:rabi_freq}
	\Omega_k(t) = \frac{e\mathcal{M}}{\hbar}\sqrt{\frac{2\mathcal{F}_0}{\varepsilon_0 c}}\mathcal{E}_k(t)
. 
\end{equation}
We use in this paper $\mathcal{M}=1.228\times 10^{-15}$~m (see Eq.~11 in \cite{amaro_2022}). 
We deliberately chose to first simulate the normalized fields $\mathcal{E}_k(t)$ and then scale them by $\overline{\mathcal{F}}$, allowing us to reuse the calculated normalized fields  for different values of $\overline{\mathcal{F}}$ and thereby minimize computation time.

The population densities of the hyperfine sublevels are obtained by numerically solving the Bloch equations of Eqs.~\eqref{eq:bloch_equations}, with the Rabi frequency of Eq.~\eqref{eq:rabi_freq} obtained from the previously calculated normalised electric fields of Eq.~\eqref{eq:dimensionless_field}. The population densities, $(\rho_{33})_k$, obtained for a simulated field, $\mathcal{E}_k$, for a set of parameters $(\overline{\mathcal{F}}, \Delta\omega, \tau, D, R)$ are averaged over $K$ for $k=1,2,...,K$ to obtain ${\overline{\rho_{33}}}(t, \Delta\omega, \overline{\mathcal{F}}, \tau, D, R)$.

The effects of Doppler broadening, caused by the motion of the $\mu$p atoms within the H$_2$ gas mixture, are included using a convolution of the average ${\overline{\rho_{33}}}$ population density with the Gaussian distribution describing the Doppler profile~\cite{amaro_2022}:
\begin{equation}
	\label{eq:doppler_conv}
	\overline{\rho_{33}}_{\text{D}}(t, \Delta\omega) = \int_{-\infty}^{+\infty}  \frac{\overline{\rho_{33}}(t, \omega^{\prime})}{\sqrt{2\pi}\sigma_{\text{D}}}\exp\left[-\frac{(\Delta\omega - \omega^{\prime})^2}{2\sigma_{\text{D}}^2}\right] d\omega^{\prime},
\end{equation}

\noindent where  $\sigma_{\text{D}} = \omega_r\sqrt{\frac{k_B T}{(m_{\mu} + m_p)c^2}}$ is the Doppler width, with $k_B$ being the Boltzmann constant, $T$  the target temperature,  $m_{\mu}$ the muon mass and $m_p$ the proton mass.

\section{Numerical results}
\label{sec:results}

\begin{figure}[h]
	\centering
	\includegraphics[width = 0.49\textwidth]{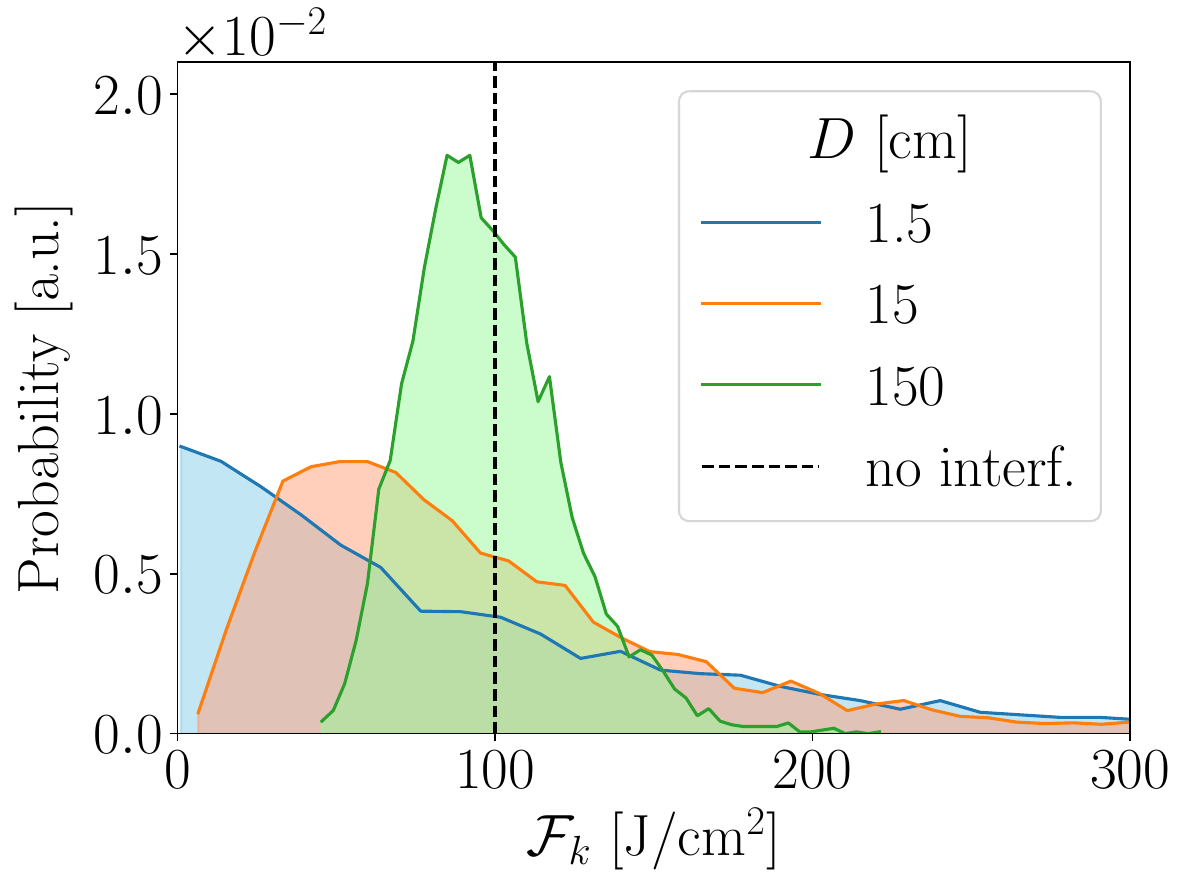}
    \hfill
    \includegraphics[width = 0.49\textwidth]{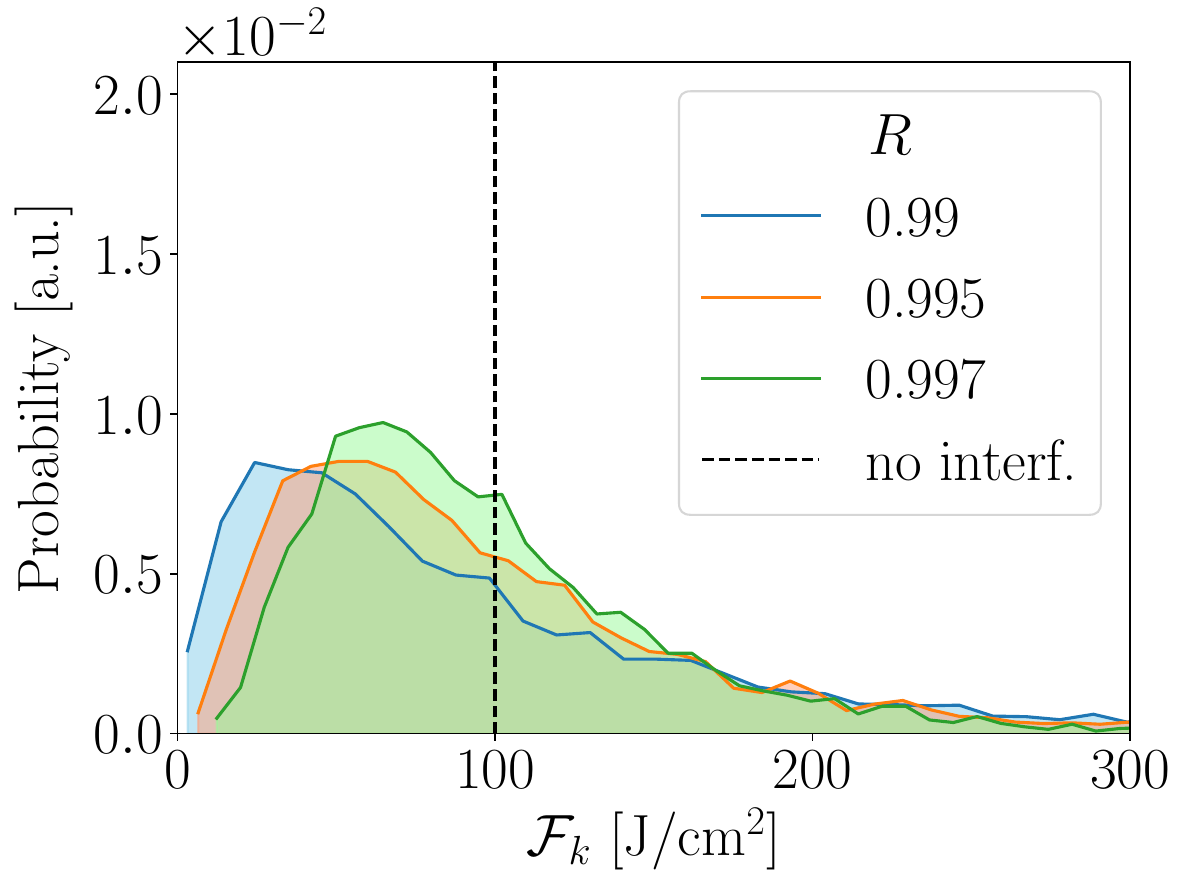}
	\caption{Fluence distributions for a fixed average fluence of $\overline{\mathcal{F}} = 100$~J/cm$^2$: (Left)  for various values of $D$ at fixed  $R = 0.995$; (Right) for various values of $R$ at fixed  $D=15$~cm. The dashed vertical line represents the delta-like fluence distribution for the case without interference. In both cases we assume $\tau=50$~ns. }
	\label{fig:fluence_distributions}
\end{figure}

Figure~\ref{fig:fluence_distributions} (Left) shows the distribution of fluences $\mathcal{F}_k$ obtained for a fixed value of the average laser fluence $\overline{\mathcal{F}}=100$~J/cm$^2$, cell reflectivity $R=0.995$, and $\tau=50$~ns, for three values of $D$. The width of the fluence distribution decreases as $D$ increases, owing to the reduced overlap between folded parts of the pulse in the cell. 

Similarly, Fig.~\ref{fig:fluence_distributions} (Right) illustrates how the fluence distribution changes with varying cell reflectivity: as the number of reflections increases, the distribution narrows around the average value due to the increased number of interfering passes.

\begin{figure}[h]
	\centering
	\includegraphics[width = \textwidth]{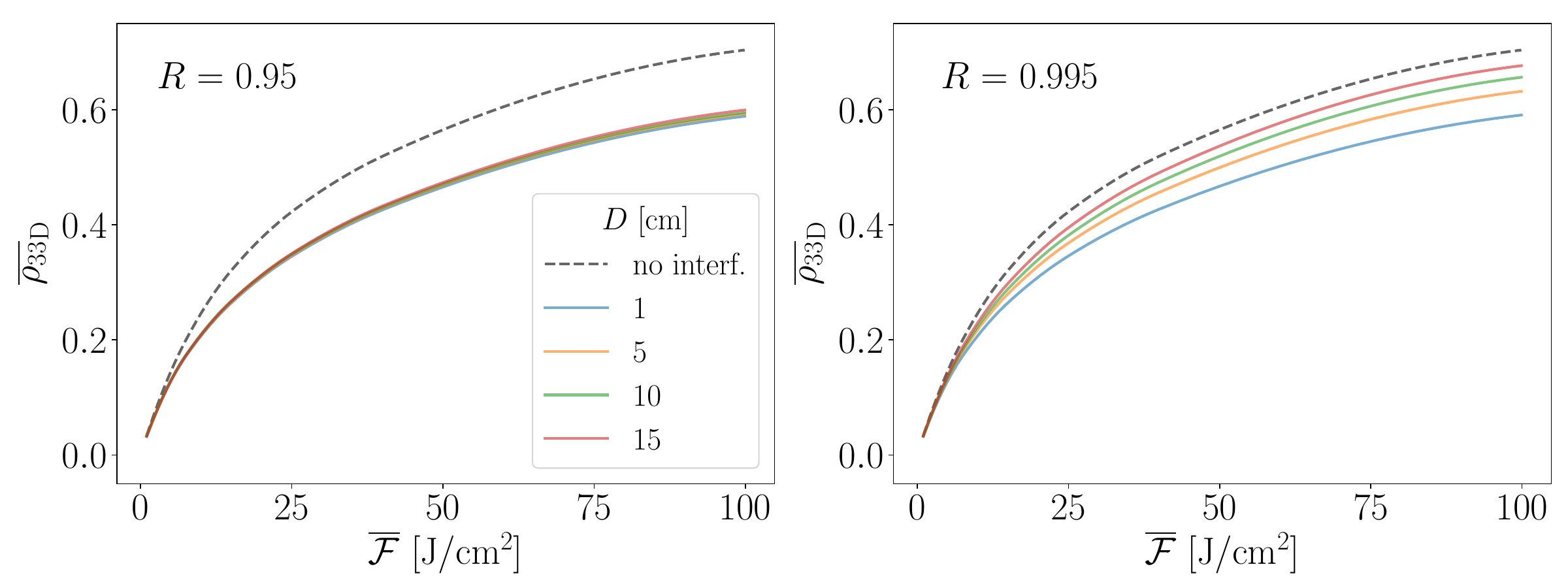}
    \includegraphics[width = \textwidth]{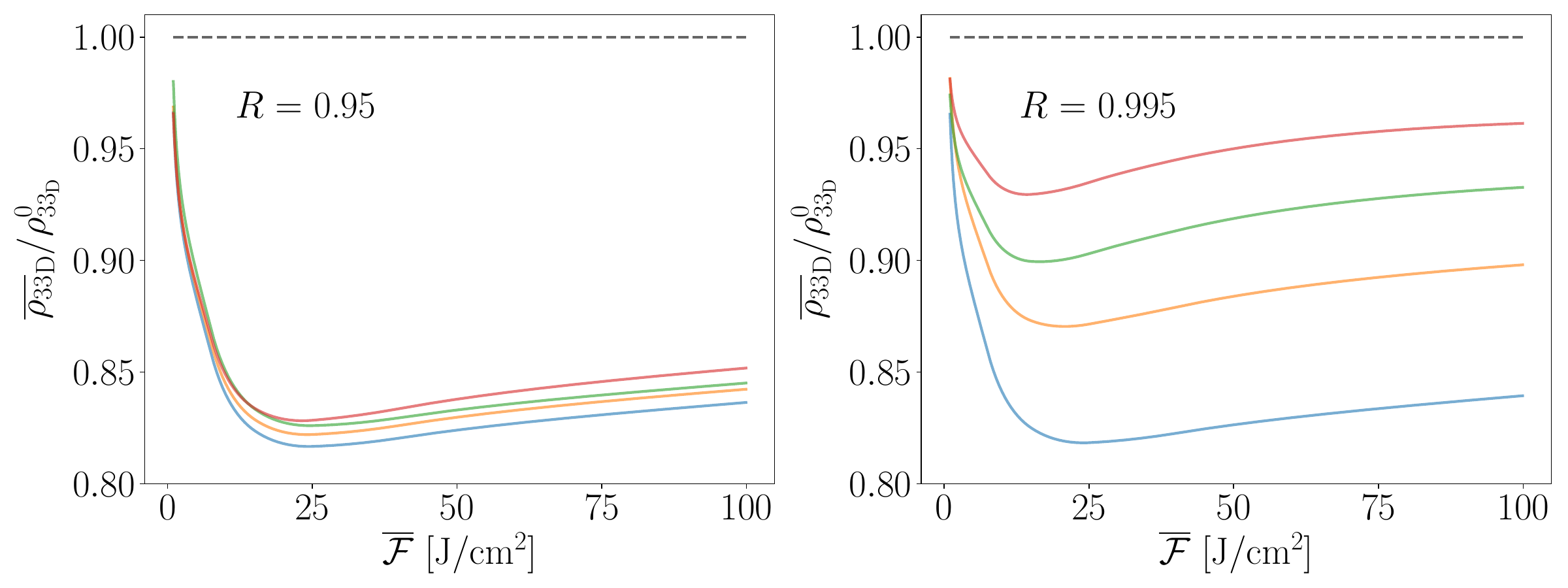}
	\caption{(Top) Simulated population density $\overline{\rho_{33}}_{\text{D}}$ (combined laser excitation followed by collisional de-excitation) versus average laser fluence $\overline{\mathcal{F}}$, for various multi-pass cell diameters, $D$, and two reflectivities, $R=0.950$ (Left) and $R=0.995$ (Right). The dashed black lines represent the case without interference effects. (Bottom) $\overline{\rho_{33}}_{\text{D}}$ population density normalized to the population density calculated neglecting interference effects, $\rho_{33_{\text{D}}}^0$. All points were calculated for $\Delta\omega=0$ (on resonance), $T=22$~K, $p=0.5$~bar, $\Delta_l=10$~MHz and $\tau=50$~ns~\cite{amaro_2022, Nuber2023}.}
	\label{fig:fluence_saturation}
\end{figure}

In Fig.~\ref{fig:fluence_saturation} (Top) we present the simulated laser-induced excitation probability (at resonance), including subsequent collisional de-excitation, $\overline{\rho_{33}}_{\text{D}}(\Delta\omega=0)$, as a function of the average fluence $\overline{\mathcal{F}}$, for two reflectivities $R$ and various cell diameters $D$.

The dashed black curves show the population density without interference effects, which we represent by $\rho_{33_{\text{D}}}^0$, versus laser fluence.
The other coloured curves have been calculated by taking into account the interference effects from the beam folding in the cell.

The transition probability decreases once interference effects are included.
Indeed, due to saturation effects—i.e. deviations from the linear dependence of $\rho_{33}$ on the average fluence—the average transition probability
$\overline{\rho_{33}(\mathcal{F}_k)}$ (averaged over $\mathcal{F}_k$)
is significantly smaller than $\rho_{33}(\overline{\mathcal{F}})$.

To better quantify the reduction in the simulated transition probability relative to the reference case without interference (dashed black curves), the bottom panels of Fig.~\ref{fig:fluence_saturation} show the ratio of the interference-inclusive result to the one neglecting interference.
As expected, the decrease in transition probability due to interference effects is negligible 
for small values of $\overline{\mathcal{F}}$, where saturation effects are basically absent, 
and also diminishes for very large $\overline{\mathcal{F}}$, where the saturation itself 
has reached its maximum.
In the intermediate region, the combined laser-excitation and collisional de-excitation probability decreases by less than $20\%$ for any fluence value.
The comparison between the left and right panels of Fig.~\ref{fig:fluence_saturation} clearly illustrates that the reduction in transition probability becomes smaller with increasing reflectivity
$R$, due to the narrowing of the $\mathcal{F}_k$ distribution shown in Fig.~\ref{fig:fluence_distributions} (Right).

 For parameters relevant to the HFS  experiment under preparation --- a cell diameter of $D = 10\ \mathrm{cm}$, a pulse length of $\tau = 50\ \mathrm{ns}$, and an effective (including  coupling slit  losses) reflectivity of $0.990 < R < 0.997$ --- the maximal reduction in transition probability is only $\lesssim 10\%$, independent of the fluence value.

\section{Conclusions}

We have presented a simple model to estimate the maximal reduction of the laser-induced transition probability due to wave-interference effects in a multi-pass cell, in the regime where the pulse length exceeds the round-trip time of the cell. The approach first quantifies maximal interference in terms of the laser electric field and the corresponding Rabi frequency, and then incorporates this into the population dynamics via the optical Bloch equations.

For the HFS experiment in muonic hydrogen, the resulting decrease in transition probability due to interference effects is found to be $\lesssim 10\%$ provided $D \gtrsim 10$~cm, $\tau \lesssim 50$~ns and $R \gtrsim 0.99$, as visible in Fig.~\ref{fig:fluence_saturation} for any fluence value. Hence, this result applies to any  spatial distribution of fluence in the multi-pass cell, which depends on various experimental conditions such as in-coupling angle, in-coupled energy, and cell reflectivity. Since the results of Fig.~\ref{fig:fluence_saturation} represent an upper bound—assuming a larger interference effect than would actually occur in the real cell—we conclude that, under the above given conditions, interference effects can be safely neglected.
Furthermore, this study offers a valuable guide for optimizing the overall performance of the experimental setup.

The method outlined here may also be of value in other contexts where coherent light in a multi-pass geometry is used to drive weak atomic or molecular transitions. The normalized plots in Fig.~\eqref{fig:fluence_saturation} (Bottom) can serve as a practical reference for assessing the potential impact of such effects in similar scenarios.

\section*{Funding}
We acknowledge the support of the following grants: Fundação para a Ciência e a Tecnologia (FCT), Portugal, with projects UID/04559/2020 (LIBPhys) and contracts No. 2023.04065.BD and 2023.01585.BD; 
European Research Council (GoG \#725039); Swiss National Science Foundation, Grant Number 197052.

\section*{Disclosure statement}
No potential conflict of interest was reported by the author(s).

\section*{Data availability statement}
The code used to produce the findings of this study is
available on the Zenodo repository with the identifier
doi:10.5281/zenodo.17202955.

\bibliographystyle{tfo}

\end{document}